\documentclass[11pt]{article}
\pdfoutput=1
\usepackage{jcappub}
\usepackage{array}
\usepackage{aas_macros}
\usepackage{graphicx}
\usepackage{epsfig}
\usepackage{subfigure}
\usepackage{mathrsfs}
\usepackage[normalem]{ulem}
\usepackage{algorithm}
\usepackage{algpseudocode}

\newcommand{\be}{\begin{equation}}
\newcommand{\ee}{\end{equation}}
\newcommand{\gpc}{\, {\rm Gpc}}
\newcommand{\mpc}{\, {\rm Mpc}}

\newcommand{\hmpc}{\, h^{-1} \mpc}
\newcommand{\hgpc}{\, h^{-1} \gpc}

\newcommand{\FastPM}{\textsf{FastPM}}

\title{High mass and halo resolution from fast low resolution simulations}  
\author[a]{Biwei Dai,}
\author[a]{Yu Feng,}
\author[a,b,c]{Uro\v s Seljak,}
\author[a]{Sukhdeep Singh}

\affiliation[a]{Berkeley Center for Cosmological Physics and Department of Physics, University of California, Berkeley, CA 94720, USA}
\affiliation[b]{Department of Astronomy, University of California, Berkeley, CA 94720, USA}
\affiliation[c]{Lawrence Berkeley National Lab, 1 Cyclotron Road, Berkeley, CA 94720, USA}

\emailAdd{biwei@berkeley.edu}
\emailAdd{yfeng1@berkeley.edu}
\emailAdd{useljak@berkeley.edu}
\emailAdd{sukhdeep1@berkeley.edu}

\abstract{Generating mocks for future sky surveys requires large volumes and high resolutions, which is computationally expensive even for fast simulations. In this work we try to develop numerical schemes to calibrate various halo and matter statistics in fast low resolution simulations compared to high resolution N-body and hydrodynamic simulations. For the halos, we improve the initial condition accuracy and develop a halo finder "relaxed-FOF", where we allow different linking length for different halo mass and velocity dispersions. We show that our relaxed-FoF halo finder improves the common statistics, such as halo bias, halo mass function, halo auto power spectrum in real space and in redshift space, cross correlation coefficient with the reference halo catalog, and halo-matter cross power spectrum. We also incorporate the potential gradient descent (PGD) method into fast simulations to improve the matter distribution at nonlinear scale. By building a lightcone output, we show that the PGD method significantly improves the weak lensing convergence tomographic power spectrum. With these improvements \FastPM{}
is comparable to the high resolution full N-body simulation of the same mass resolution, with two orders of magnitude 
fewer time steps. These techniques can be used to improve the halo and matter statistics of \FastPM{} simulations for mock catalogs of future surveys such as DESI and LSST.}

\begin{document}
\maketitle
\flushbottom

\section{Introduction}

Numerical simulations of large scale structure formation are essential for extracting cosmological information from current and future sky surveys. N-body simulations with semi-analytic galaxy formation models 
have achieved great success in cosmological analysis \cite{kitaura2016a, joudaki2018a, maccrann2018a}, but they are also computationally expensive.

Quasi N-body PM simulations with a small number of steps such as \FastPM{} \cite{feng2016a} and COLA \cite{tassev2013a} provide an alternative and fast way to model galaxy statistics. It has been shown that these fast simulations predict accurate halo statistics compared to full N-body simulations of the same resolution \cite{feng2016a, tassev2013a}. However, to generate accurate mocks for future sky surveys such as DESI \cite{levi2013a} and LSST \cite{lsst2009a}, high mass resolution and large box volumes are needed, which makes the computational cost quite high even for fast simulations. For example, DESI aims at measuring the bright emission line galaxies up to $z=1.7$, the analysis of which requires accurate modeling of $10^{11}M_{\odot}$ halos \cite{desi2016a}. Considering that using halos with less than 200 particles could lead to large systematic errors \cite{derose2019a}, and that to cover the sky up to $z=1.7$ the box should be around $3\hgpc$ per side, we need at least 6 trillion dark matter particles in the simulation. This is computationally expensive in itself even with fast simulations like \FastPM{}, not to mention that we may need lots of different realizations to measure the covariance matrices or to study the influence of cosmological parameters. Therefore, we need to find a model that reduces the computation cost while maintaining the accuracy.

Another difficulty in these quasi N-body simulations is the deficiency of matter power on small scale. The potential gradient descent (PGD) model has been proposed to improve the modeling of matter distribution on nonlinear scales \cite{dai2018a}. PGD was used as a post processing correction on the static snapshot, while here we want to further incorporate PGD into \FastPM{} per time step, so that it could be used in generating time-continuous light-cone mocks for weak lensing analysis.


The goal of this paper is to produce reliable predictions for halo and dark matter statistics in low resolution \FastPM{} simulations by training them on high 
resolution N-body simulations. The plan of the paper is as following. In section \ref{sec:halo} we try to improve the identification of small halos by modifying the FoF halo finder and removing fake halos. For the matter field, we incorporate PGD into \FastPM{} simulation in section \ref{sec:matter}. By building a light-cone output we show that the method can improve the weak lensing convergence field. Finally we conclude in Section \ref{sec:conclusion}.

\section{Halo statistics and clustering}
\label{sec:halo}

In this section we examine and improve the halo statistics in \FastPM{} simulation. We focus on halos larger than $10^{11} M_{\odot}$, and use IllustrisTNG \cite{springel2018a} as our reference simulation. IllustrisTNG is a suite of cosmological hydrodynamic simulations with different box sizes and resolutions. We will mainly compare our results with TNG300-2-Dark, a dark-matter-only run in a $205 \hmpc$ periodic box and with $1250^3$ particles. Since previous study shows that halo statistics have around $2\%$ deviations for halos consisting of 200 particles \cite{derose2019a}, TNG300-2-Dark may not be accurate enough for the halo mass we consider ($10^{11} M_{\odot}$ halo consists of 180 particles), so we also examine the TNG300-1-Dark simulation, which has a 8 times higher resolution. The agreement between TNG300-2-Dark and TNG300-1-Dark should give us an estimate of the accuracy of our reference simulation. Besides, we also show the results from TNG300-2 hydrodynamic simulation to study the baryonic effects on these halo statistics. All these simulations share the same initial linear density field.

To perform a direct comparison with the reference simulation, we run \FastPM{} in the same $205 \hmpc$ periodic box with the same linear density field by matching the random seed and linear power spectrum. We generate the initial condition at $z=9$ using 2LPT, and then evolve the field to redshift 0 with 40 steps distributed uniformly on the scale factor $a$. Unlike TNG300-2-Dark with $1250^3$ dark matter particles, we run \FastPM{} simulation with 8 times lower resolution, i.e., $625^3$ particles. For most comparisons in this paper the halo catalogs are generated from abundance matching, and the smallest halo in \FastPM{} halo catalogs has around 20 particles. In the following analysis, we also compare our results with TNG300-3-Dark, which has the same resolution but is a full N-body simulation. As we will see, with standard linking length $l=0.2$, most of the comparisons with TNG300-3-Dark is quite good, consistent with previous study \cite{feng2016a}, but the agreement is not so satisfying when compared to the higher resolution reference simulation. 
This suggests that halos with less than a hundred particles cannot be modeled well even with a full N-body simulation, and the deficiency of \FastPM{} at this mass range is mostly a resolution issue. This provides 
an additional motivation to our 
approach: by training on 
Illustris TNG300-2-Dark, which has a higher overall resolution (mass, time, and force),
we can obtain results with \FastPM{} that 
can exceed even Illustris TNG300-3-Dark despite its 
higher time and force resolution. We do so 
by modifying the standard Friends-of-Friends (FoF) algorithm to improve the situation for these small halos. All the halos in the Illustris TNG simulations are identified using FoF algorithm with linking length 0.2.

\subsection{Relaxed-FoF}

\begin{figure}
\includegraphics[width=\textwidth]{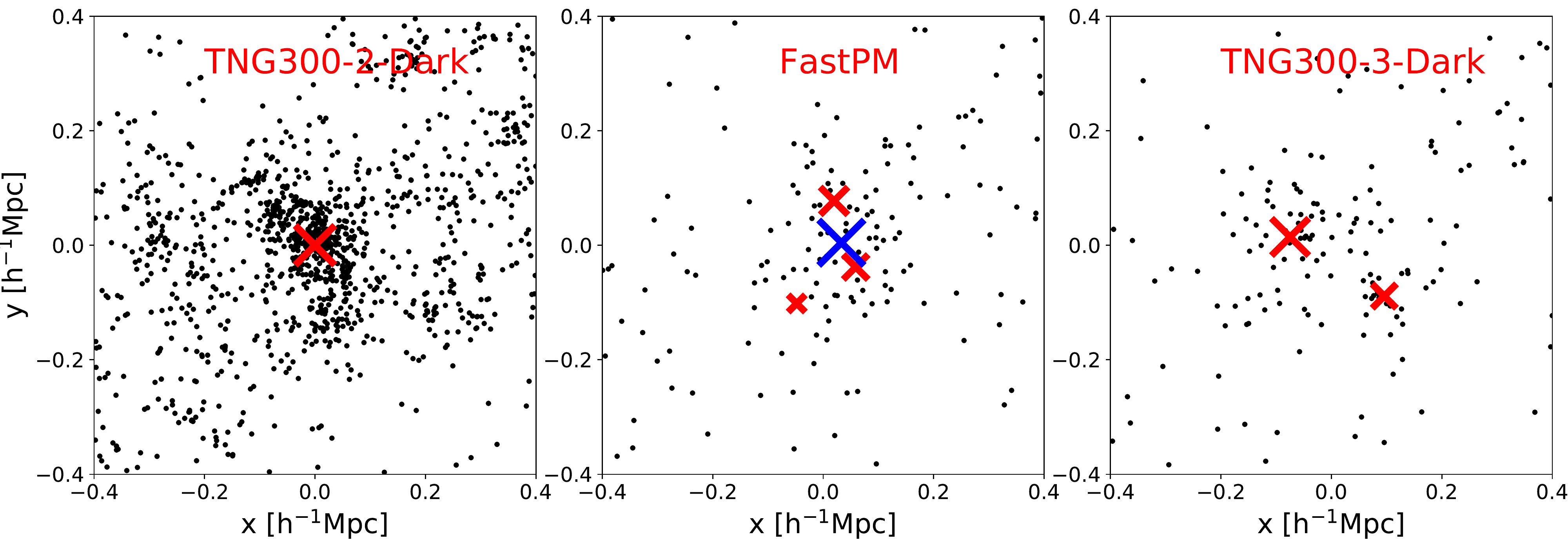}
\caption{
The projected map of the same halo in TNG300-2-Dark (left panel), \FastPM{} (middle panel) and TNG300-3-Dark (right panel). The halo mass is around $2.4 \times 10^{11} M_{\odot}$, corresponding to 425 particles in TNG300-2-Dark simulation, and 53 particles in \FastPM{} and TNG300-3-Dark. We perform the standard FoF algorithm with linking length 0.2 on each of them, and the halo centers of mass are represented as red crosses (the size of cross is proportional to the halo mass). The blue cross in the middle panel shows the center of mass of the halo identified with a larger linking length.}
\label{fig:ProjectedHalo}
\end{figure}

\begin{figure}
\includegraphics[width=\textwidth]{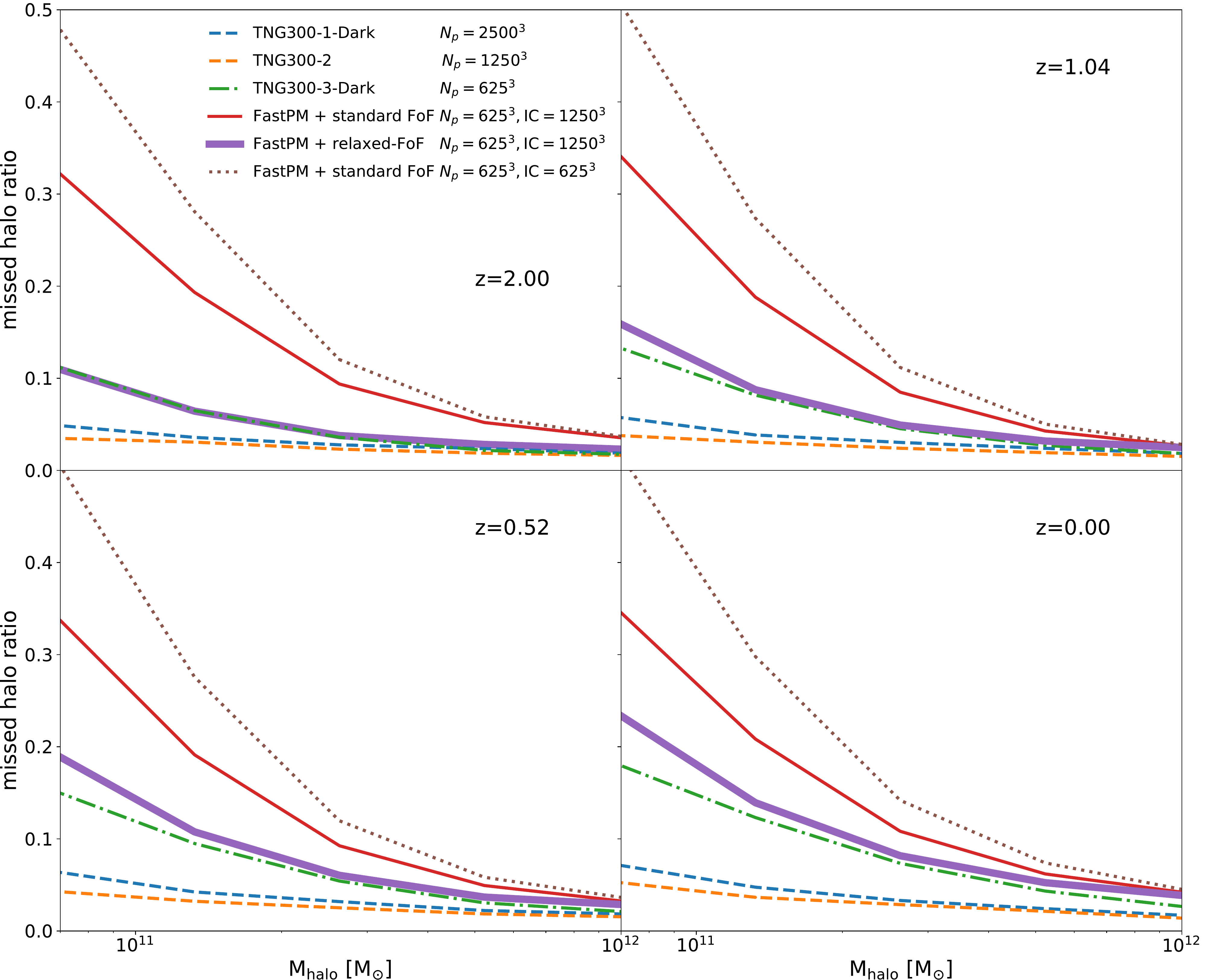}
\caption{The ratio of missed halo 
in different simulations and various redshifts. Here we choose TNG300-2-Dark as the reference (same for Figure \ref{fig:HaloBias} to Figure \ref{fig:CC_halo-matter}). The nonzero ratio of missed halo in the massive end is mostly because of the bridging effect.
}
\label{fig:MissedHaloRatio}
\end{figure}

Large halos in \FastPM{} can be modeled accurately, but small halos cannot be well resolved. For example, in Figure \ref{fig:ProjectedHalo} we show the same halo in the high resolution reference simulation, \FastPM{}, and a full N-body simulation with the same resolution TNG300-3-Dark. We see that the lower mass resolution halo is more diffused than the reference, the FoF algorithm with standard 0.2 linking length cannot link all the particles, therefore breaking the halo into 2 or 3 smaller halos. As a result, \FastPM{} and TNG300-3-Dark tend to underestimate the halo mass at this mass range, and therefore underestimate the mass function, as we can see in Figure \ref{fig:HaloMassFunction}. Because halos are broken into several small halos, they appear to be more clustered and produce a larger halo bias (Figure \ref{fig:HaloBias}). 

\begin{algorithm}
\caption{Relaxed-FoF algorithm\label{alg:relaxedFoF}}
\begin{algorithmic}[1]
\Procedure {RelaxedFoF}{$x$, $N_p$, $l$, $r$} \Comment{\parbox[t]{.45\linewidth}{$x$ is the set of particles, $N_p$ is the halo bin in ascending order, $l$ is the corresponding linking length, and $r$ is the velocity dispersion threshold.}}
\For {$i \leftarrow 1, \mathrm{len}(N_p)$} 
\State $\mathrm{halo} \leftarrow$ FoF($x, l[i]$) 
\For {$j \leftarrow 1, N_{\mathrm{halo}}$} 
\If {$\mathrm{halo[j].Np} < Np[i]\ \mathrm{\bf and}\ \mathrm{halo[j]}.V_{\mathrm{disp}} < r~V_{\mathrm{std,disp}}(\mathrm{halo[j].mass})$}
\State save halo[j] in halocat
\State remove particles that form halo[j] from $x$
\EndIf
\EndFor
\State remove particles that do not form halos from $x$
\EndFor
\\
\State $L \leftarrow l[\mathrm{len}(N_p)]$
\While {$x$ is not empty} \Comment{\parbox[t]{.45\linewidth}{keep reducing the linking length and save true halos \& reject fake halos until no halos can be found}}
\State $L \leftarrow 0.9L$
\State $\mathrm{halo} \leftarrow$ FoF($x, L$)
\For {$i \leftarrow 1, N_{\mathrm{halo}}$} 
\If {$\mathrm{halo[j]}.V_{\mathrm{disp}} < r~ V_{\mathrm{std,disp}}(\mathrm{halo[j].mass})$}
\State save halo[j] in halocat
\State remove particles that form halo[j] from $x$
\EndIf
\EndFor
\State remove particles that do not form halos from $x$
\EndWhile
\EndProcedure
\end{algorithmic}
\end{algorithm}

We try to improve this situation by increasing the linking length $l$ in the FoF halo finder. In the middle panel of Figure \ref{fig:ProjectedHalo}, we see that with a larger linking length, we can successfully link all the particles and reproduce the correct halo mass and position. However, increasing the linking length for all the particles will bias the halo mass for large halos, since we know that the standard linking length $l=0.2$ is already good for large halos at redshift 0 \cite{feng2016a}. Therefore, we make the linking length a function of the halo particle number, with larger linking length for smaller halos.  Since the linking length is not fixed, we call this method relaxed-FoF. As is shown in Figure \ref{fig:HaloMassFunction}, the 0.2 linking length predict less massive halos at high redshifts, suggesting that the linking length should also be a function of redshift as well. 

Another issue with these low resolution simulations is that in the high density region, unbound clusters of nearby particles are linked with FoF algorithm, therefore producing lots of fake halos. With longer linking lengths, we expect this issue to be more severe. After some analysis on these fake halos, we find that they are likely to have larger velocity dispersion. This is expected, since the particles that make up those fake halos are just close "by accident" and are not gravitationally bounded. Therefore, for each small halo we calculate the quantity $r=\frac{V_{disp}}{V_{std, disp}(M, z)}$, where $V_{disp}$ is the velocity dispersion we measured from simulation, and $V_{std, disp}(M, z)$ is the velocity dispersion predicted by its scaling relation with the halo mass \cite{evrard2008a}.
\be
V_{std, disp}(M, z) = V_0\ \left(\frac{E(z)M}{10^{15} h^{-1}M_{\odot}}\right)^{1/3}
\ee
where $V_0\simeq 1100 \mathrm{km s^{-1}}$, $E(z)=H(z)/H(0)$ is the dimensionless hubble parameter. If the quantity $r$ is larger than a threshold $r_{0}$, we consider the halo as a fake one and reject it from the halo catalog. Since the fake halos we remove are mostly in high density region, we expect this procedure to reduce the bias of small halos. 

We divide the halos into several bins $N_{p,i}$ ($N_{p,i}$ is the maximum halo particle number of bin $i$), for each bin we have the corresponding linking length $l(N_{p,i}, z)$. We first run the FoF halo finder on all the particles in the snapshot with linking length $l(N_{p,1}, z)$ for the smallest halo bin. Then we select all the halos that are larger than the halo particle number $N_{p,1}$ and the halos that are rejected by the velocity dispersion criterion, and rerun the FoF halo finder on the particles that form these halos with the linking length $l(N_{p,2}, z)$ of the next bin. We repeat this procedure until we finish the largest halo bin. For the rest of the particles that form the fake halos, we keep running the FoF halo finder and reducing the linking length, with fake halos rejected at each iteration, until there are no particles left. The function $l(N_{p,i}, z)$ and $r_0(z)$ are simple functions we choose to produce correct halo mass function and halo bias.

\begin{align}
N_{p,i} &= \{20, 40, 80, 160, 320, \inf\}\\
l(N_{p,1}, z) &= l_1-\frac{A_1}{1+z}\\
l(N_{p,6}, z) &= \max(l_6-\frac{A_2}{1+z},\ 0.2)\\
l(N_{p,i}, z) &= \frac{(6-i)N_{p,1} + (i-1)N_{p,6}}{5}
\end{align}
\be
r_0(z) = B_1 - B_2\log(1+z)
\ee
where $l_1$, $l_6$, $A_1$, $A_2$, $B_1$ and $B_2$ are free parameters.


In addition to improving the halo finder algorithm, we also find that the small scale power in the initial condition is crucial for the identification of small halos. We find it necessary to generate the linear density map with a mesh that is twice finer than the particle grid, which helps to improve the various halo statistics (Figure \ref{fig:MissedHaloRatio} to Figure \ref{fig:CC_halo-matter}). 

Before we examine any halo statistics in the next subsection, we first take a look at how well each individual halo can be reproduced. If two halos from two simulations are within $0.4 \hmpc$ and if their mass are within a factor of 2, then we say they are the same halo, and each halo cannot be matched with more than one halo. In Figure \ref{fig:MissedHaloRatio} we show the ratio of missed halo as a function of halo mass for different redshifts. We define missed halos as the halos that cannot find a counterpart in the other simulation. We see that as we go to smaller halos, the ratio of missed halo increases and reaches around $25\%$ ($40\%$ if the linear density map has the same resolution as particle grid) at $10^{11} M_{\odot}$ halos for \FastPM{} with constant 0.2 linking length. After switching to our relaxed-FoF, the ratio of missed halo decreases at all redshifts and all halo masses. The improvement is larger at higher redshift. In particular, the ratio of missed halo is reduced to less than $10\%$ for $10^{11} M_{\odot}$ halo at redshift 2, comparable to the full N-body simulation TNG300-3-Dark. 

In Figure \ref{fig:MissedHaloRatio} we also show the ratio of missed halo for higher resolution N-body simulation TNG300-1-Dark and hydro simulation TNG300-2, and it is surprising that the ratio is not 0. In fact, the ratio is not zero even for $10^{13} M_{\odot}$ halos, and it is mostly because of the bridging effect. If two nearby halos are linked together in one simulation, while they are identified as two separate halos in another simulation, they will not be matched using our algorithm and therefore produce a nonzero ratio of missed halo.

\subsection{Halo Statistics}

\begin{figure}
\includegraphics[width=\textwidth]{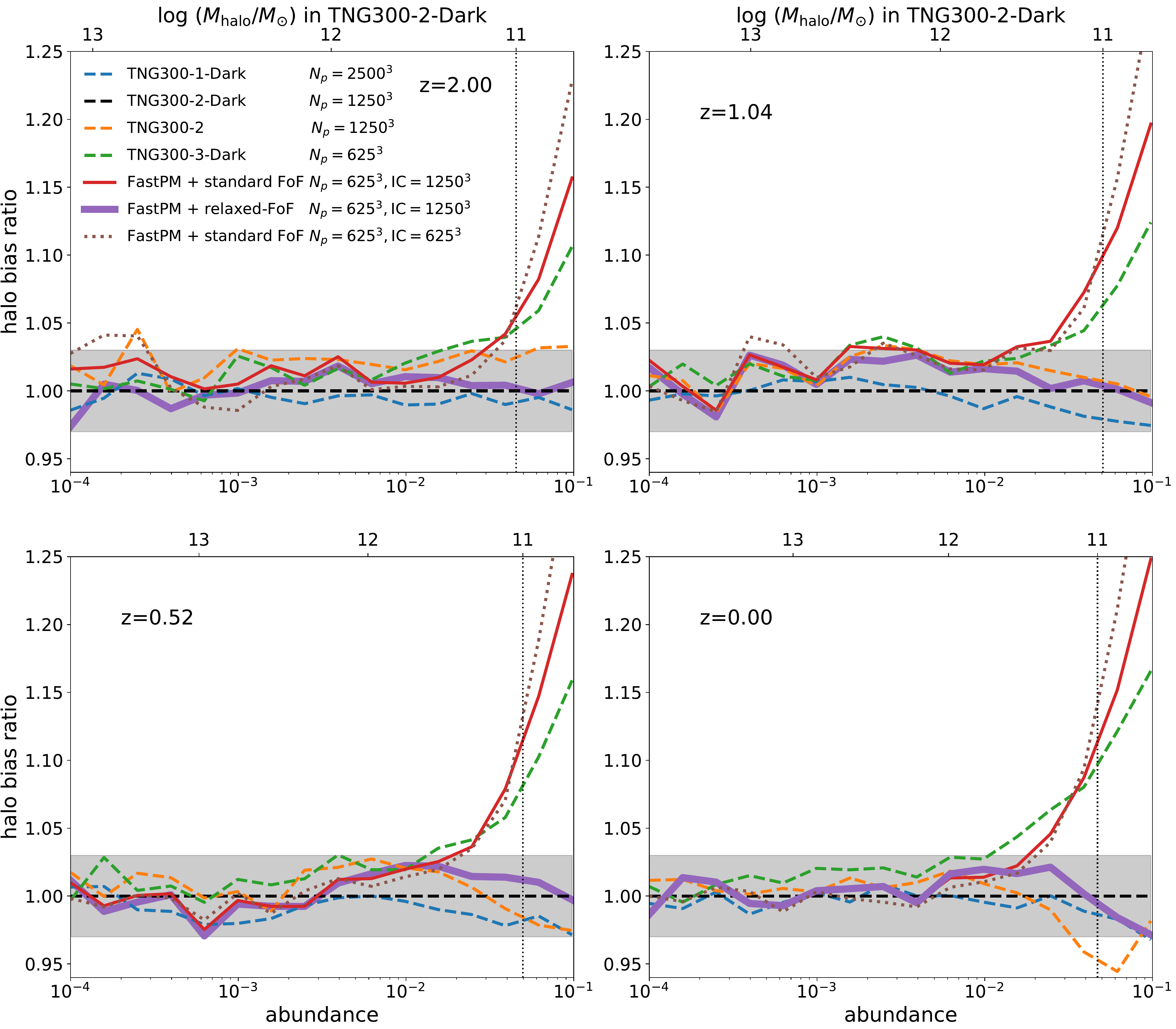}
\caption{The ratio of halo bias as a function of abundance measured in different simulations and various redshifts. The halos are selected by abundance matching, and the x axis shows the halo abundance. Larger abundance means smaller halos. The dotted vertical line shows the abundance of $10^{11} M_{\odot}$ halo, corresponding to 22 particles. The shaded region represent $3\%$ deviation. The power spectrum is calculated using Nbodykit throughout the paper \citep{nbodykit}. 
}
\label{fig:HaloBias}
\end{figure}

\begin{figure}
\includegraphics[width=\textwidth]{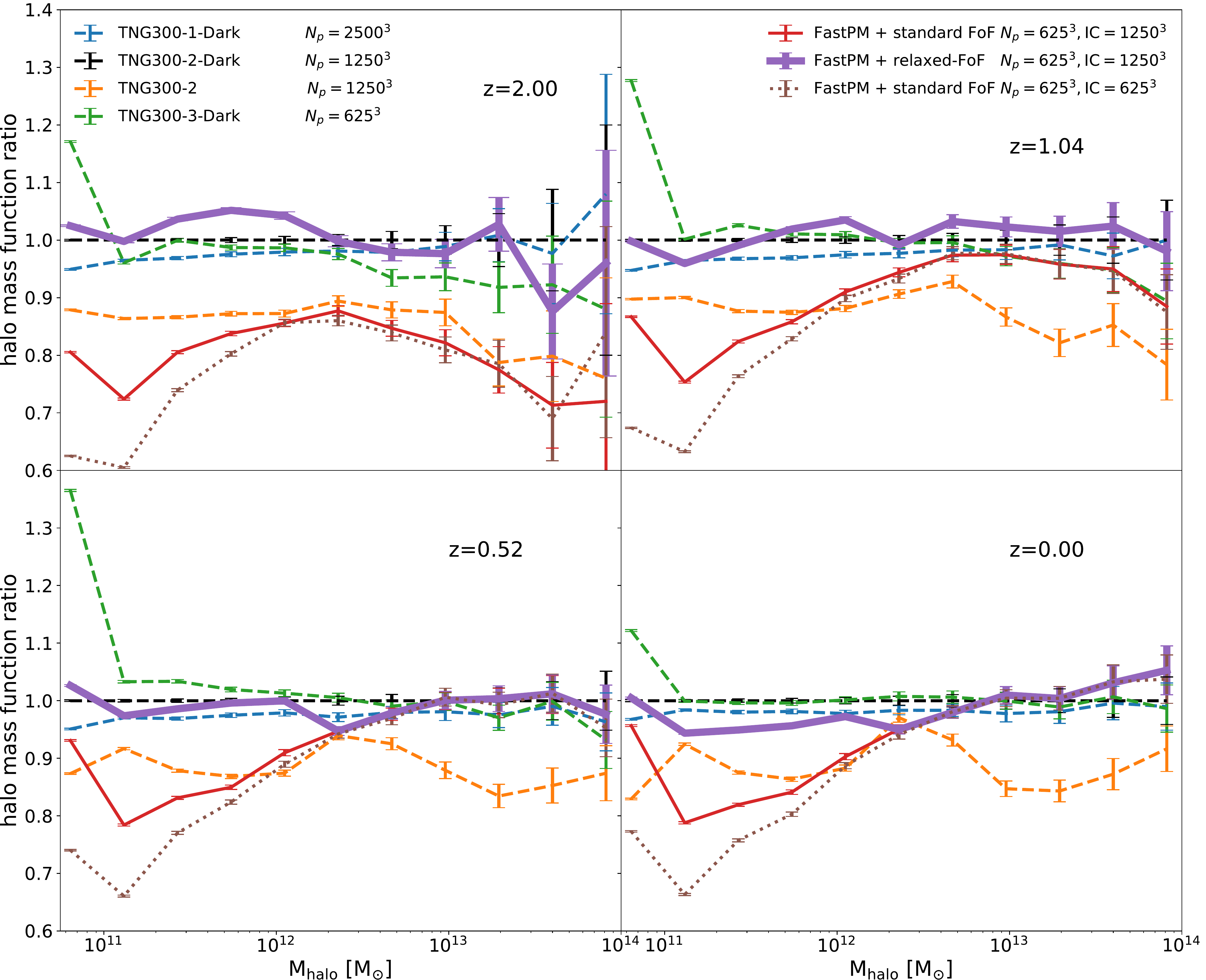}
\caption{
The ratio of halo mass function from different simulations and various redshifts. The halo mass here is defined as the FoF mass. For hydro simulation the FoF algorithm is run on the dark matter particles, and baryon particles are attached to the same groups as their nearest dark matter particle. Poisson noise errors are shown.}
\label{fig:HaloMassFunction}
\end{figure}

We first examine the halo bias defined with the halo-matter cross correlation
\be
b = \lim_{k \to 0} \frac{P_{hm}(k)}{P_{mm}(k)}
\ee
We present the halo bias results in Figure \ref{fig:HaloBias}. We see that the bias given by different simulations fluctuate a lot even for the largest halos (lowest abundance). This is because the halo mass is scattered in different simulations so the same abundance does not guarantee the same halo catalog. However, comparing the three N-body simulations of different resolutions, we can see a tendency that higher resolution simulation shows a lower halo bias, especially for small halos. 
Similarly, \FastPM{} also gives a very high bias for small halos, mostly due to its low resolution, but with our relaxed-FoF halo finder the halo bias is brought down to the normal level. 

With larger linking length, the small halos can be better identified, so the improvement of the halo mass function at the low mass end is expected (shown in Figure \ref{fig:HaloMassFunction}). With the same linking length 0.2, \FastPM{} shows a large discrepancy with full N-body simulations of all resolutions, suggesting that this deficiency in halo mass function is not due to the resolution effect, but a failure of the FoF with standard linking length to resolve small halos in \FastPM{}, 
which relaxed-FOF corrects for. Another interesting feature is that the baryonic feedback seems to reduce the mass function by $10\%$ to $20\%$, but this comparison is based on FoF mass and it is unclear if it is a meaningful comparison against hydrodynamic simulations. 

\begin{figure}
\includegraphics[width=\textwidth]{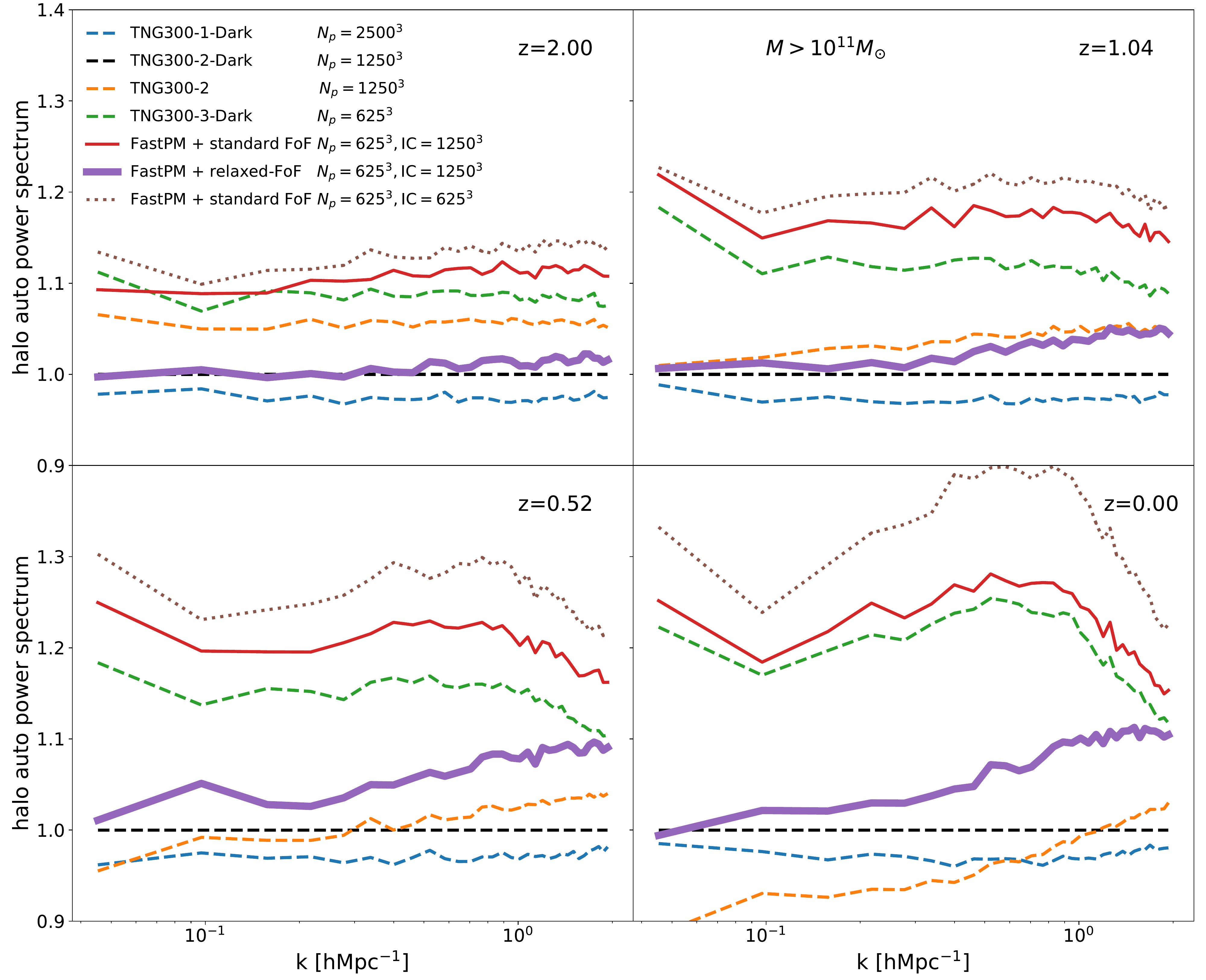}
\caption{The ratio of halo auto power spectrum from different simulations and various redshifts. Similar to Figure \ref{fig:MissedHaloRatio}, TNG300-2-Dark is chosen as our reference.
}
\label{fig:PS_halo}
\end{figure}

\begin{figure}
\includegraphics[width=\textwidth]{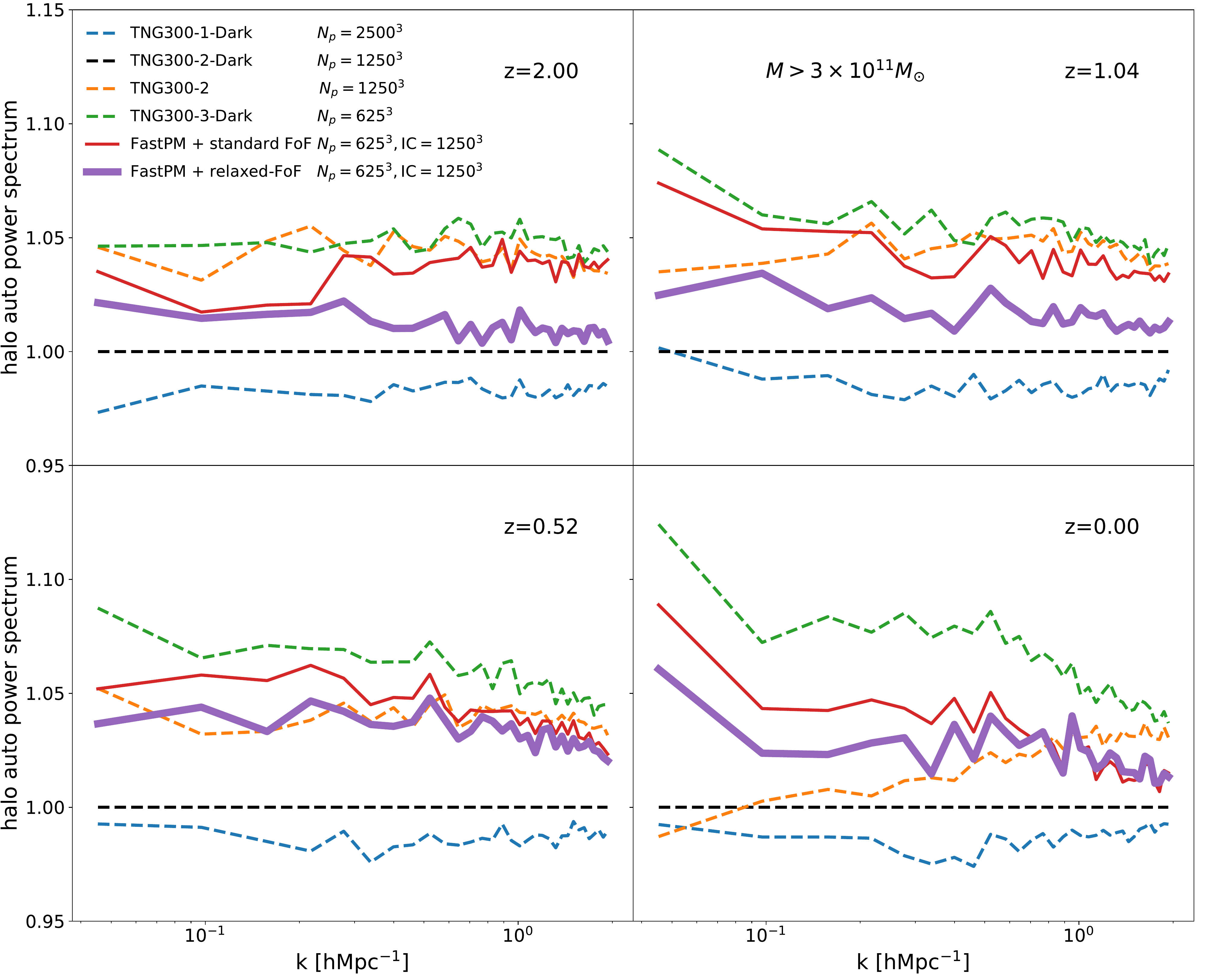}
\caption{The same as Figure \ref{fig:PS_halo}, but for halos larger than $3\times10^{11} M_{\odot}$.}
\label{fig:PS_halo1}
\end{figure}

\begin{figure}
\includegraphics[width=\textwidth]{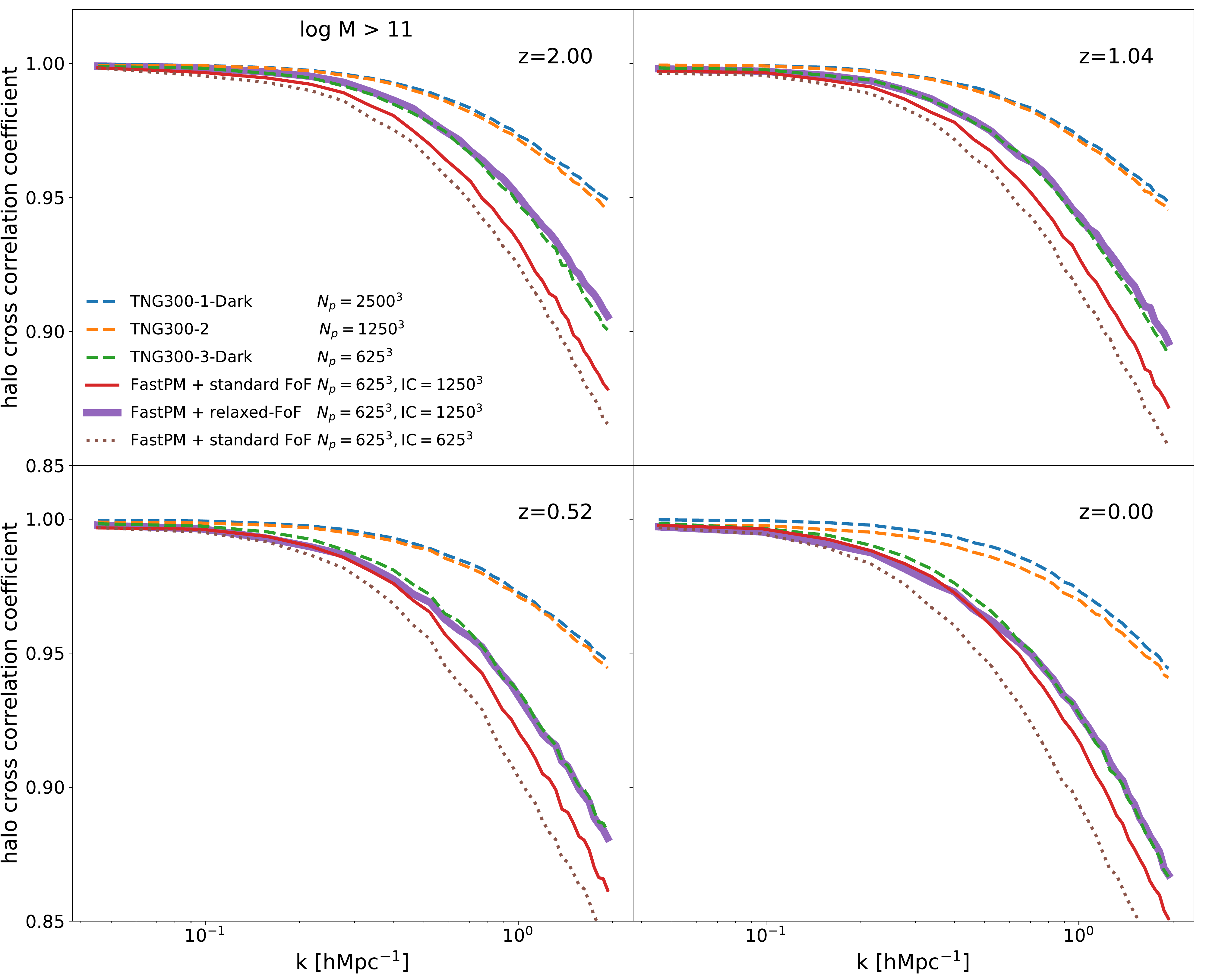}
\caption{The cross correlation coefficient of the reference halo with halos from other simulations in different redshifts.}
\label{fig:CC_halo}
\end{figure}

\begin{figure}
\includegraphics[width=\textwidth]{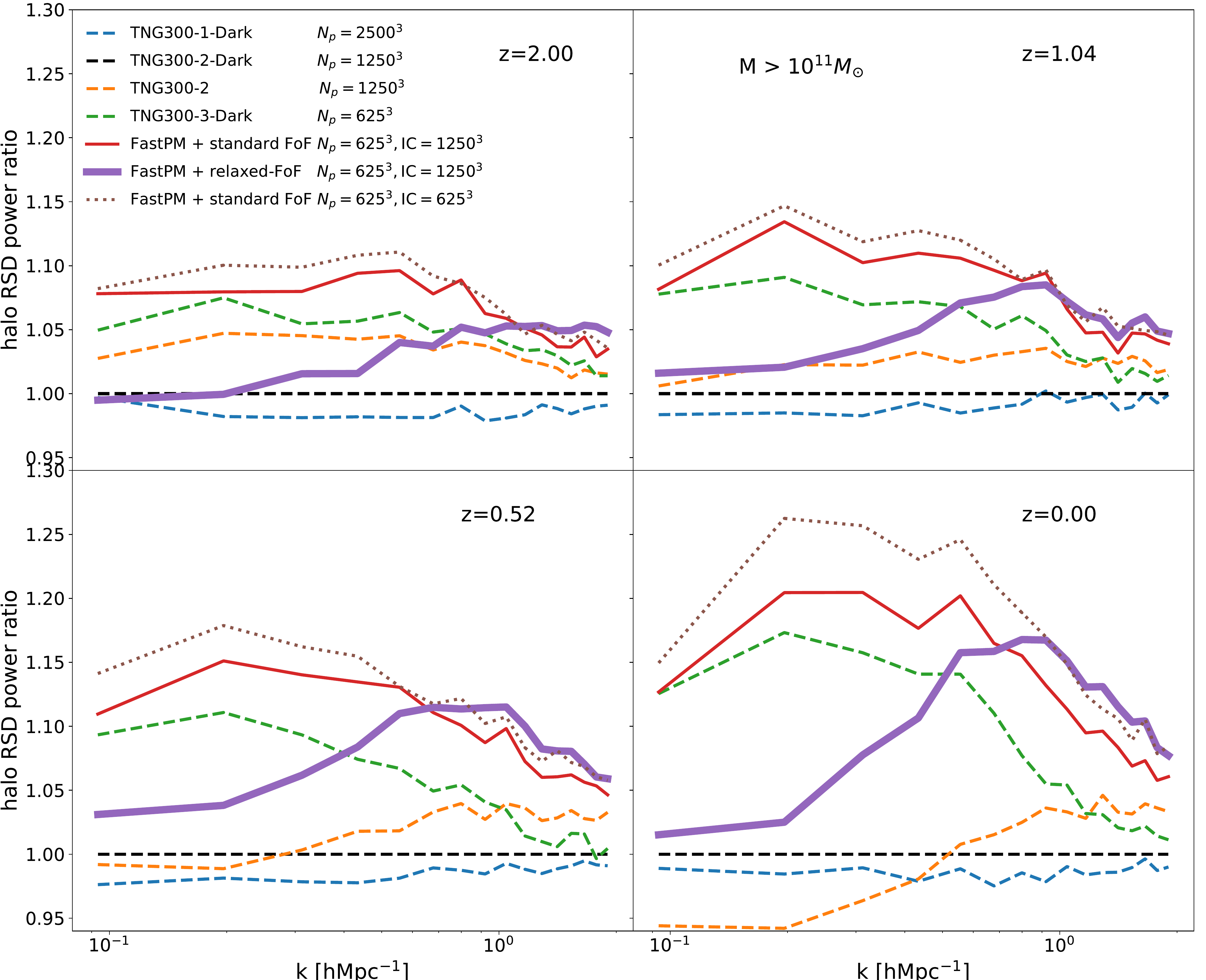}
\caption{The ratio of halo power spectrum in redshift space from different simulations and various redshifts. Here we only show the power spectrum of $k$ parallel to the line of sight, i.e., $\mu=0.9$, where $\mu=k_{\parallel}/k$. The $k$ mode perpendicular to the line of sight is not affected by RSD, and therefore the RSD halo power spectrum with $\mu=0$ is similar to the halo power spectrum in real space presented in Figure \ref{fig:PS_halo}. }
\label{fig:RSD_halo}
\end{figure}

\begin{figure}
\includegraphics[width=\textwidth]{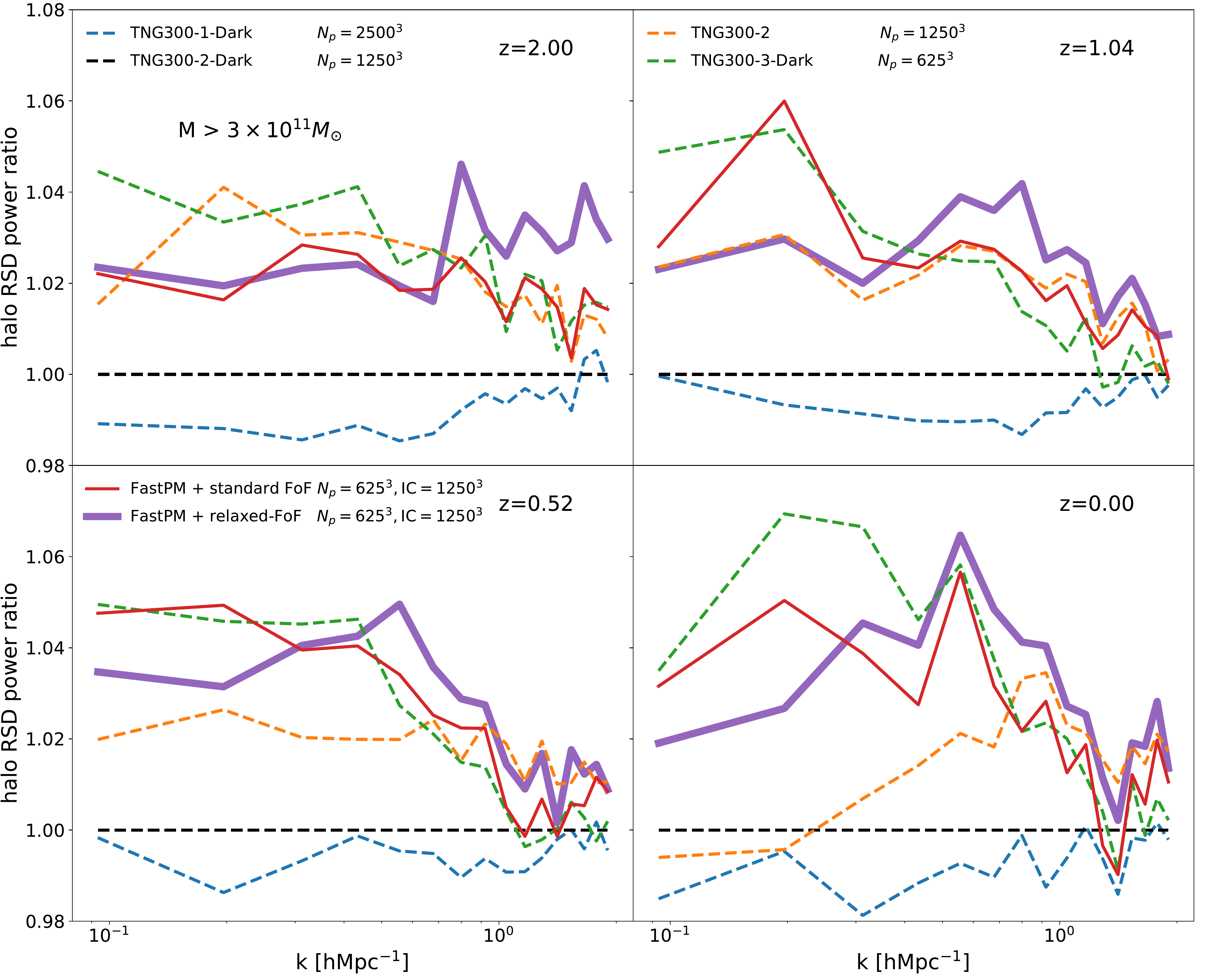}
\caption{The same as Figure \ref{fig:RSD_halo}, but for halos larger than $3\times10^{11} M_{\odot}$. For RSD halo power spectrum with $\mu=0$ see Figure \ref{fig:PS_halo1}.}
\label{fig:RSD_halo1}
\end{figure}

\begin{figure}
\includegraphics[width=\textwidth]{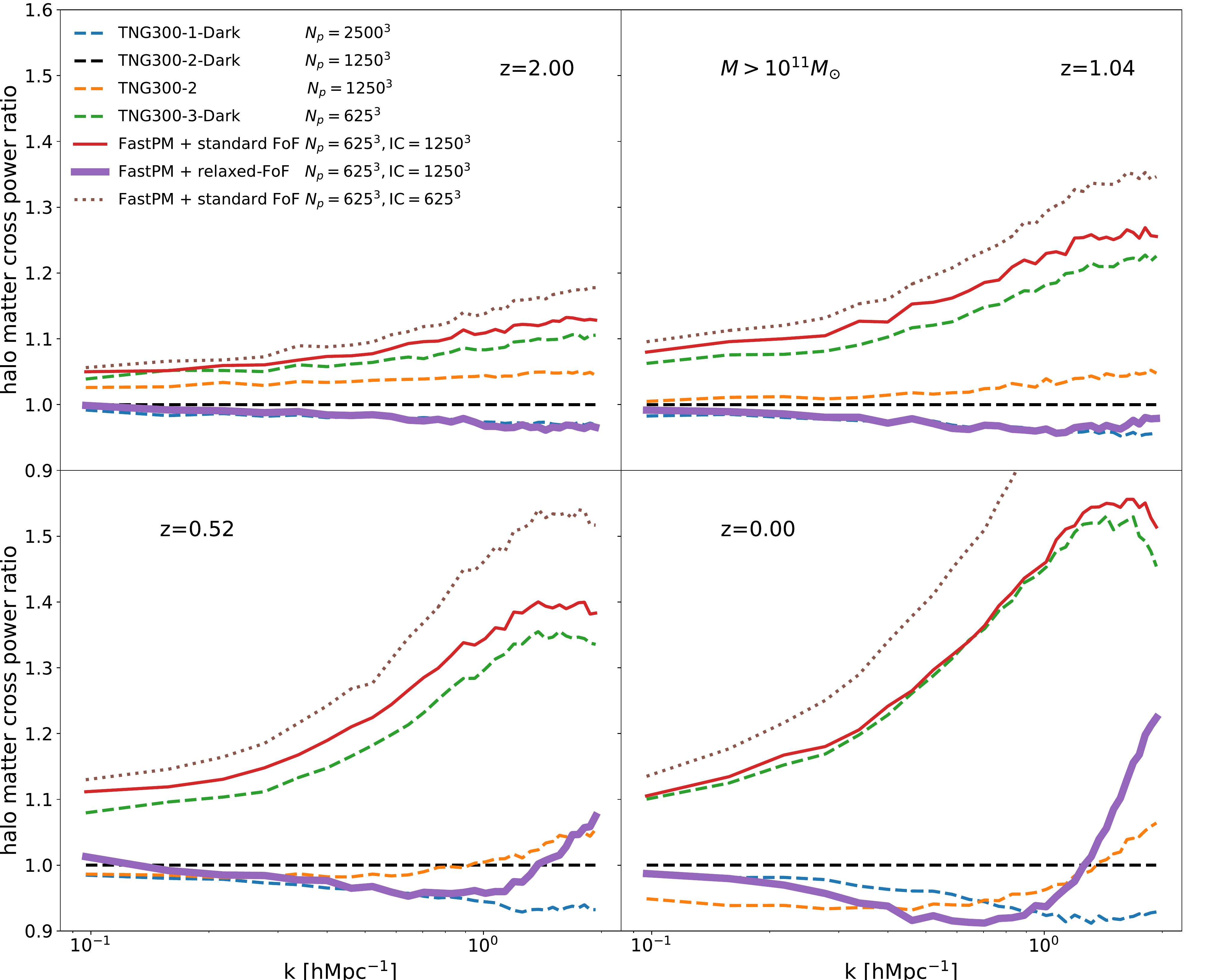}
\caption{The ratio of halo-matter cross power spectrum from different simulations and various redshifts.}
\label{fig:CC_halo-matter}
\end{figure}

Next we select halo catalogs from different simulations with abundance matching, and examine their auto power spectrum (Figure \ref{fig:PS_halo}), cross correlation coefficient with the reference simulation TNG300-2-Dark (Figure \ref{fig:CC_halo}), power spectrum in redshift space (Figure \ref{fig:RSD_halo} and Figure \ref{fig:RSD_halo1}), and the halo-matter cross power spectrum (Figure \ref{fig:CC_halo-matter}). The halo catalogs corresponding to $M \geq 10^{11} M_{\odot}$ halos. As mentioned above, the same abundance does not guarantee same halos, so we expect to see a little scatter across different simulations. As long as the deviation of \FastPM{} is comparable to the scatter of TNG-1-Dark or TNG-2, we can say the predictions of \FastPM{} is good enough. 

We see that with standard $l=0.2$ linking length, the auto power spectrum in real space and redshift space, and the halo-matter cross power spectrum of \FastPM{} are very similar (slightly worse) to those of the full N-body simulation with the same resolution. After changing to  relaxed-FoF, the halo auto power spectrum and halo-matter cross power spectrum is quite good at high redshift ($z \geq 1$), while at low redshift ($z\leq 0.5$) on small scale \FastPM{} is overpredicting power. Similar thing happens in the halo RSD power spectrum, where on large scale \FastPM{} is correct, but on smaller scale ($k \geq 0.2 \hmpc$) the power is too much. The cross correlation coefficient is consistent with the ratio of missed halo (Figure \ref{fig:MissedHaloRatio}), that after improving the halo finder the halo catalog from \FastPM{} is similar to a full N-body simulation TNG300-3-Dark.

Here in this section we only show the results of $M>10^{11}M_{\odot}$ halos. If we look at larger halos, the agreement with the reference is a lot better. For example, for $M>3\times10^{11}M_{\odot}$ \FastPM{} halos, the deviation of halo RSD power spectrum (Figure \ref{fig:RSD_halo1}) and halo-matter cross power spectrum is within $6\%$, for all scales up to $k=2\hmpc$ and all redshifts up to $z=2$.

\section{Dark Matter statistics}
\label{sec:matter}

In this section we focus on improving the matter distribution on small scale. We first incorporate PGD model into every step of \FastPM{}, and show that the redshift evolution of the PGD parameters can be parametrized by simple analytical functions. Then we build a light-cone simulation with the output of \FastPM{}, and show that the PGD model can improve the weak lensing convergence power spectrum.

\subsection{PGD Embedded in \FastPM{}}

\begin{figure}
\includegraphics[width=\textwidth]{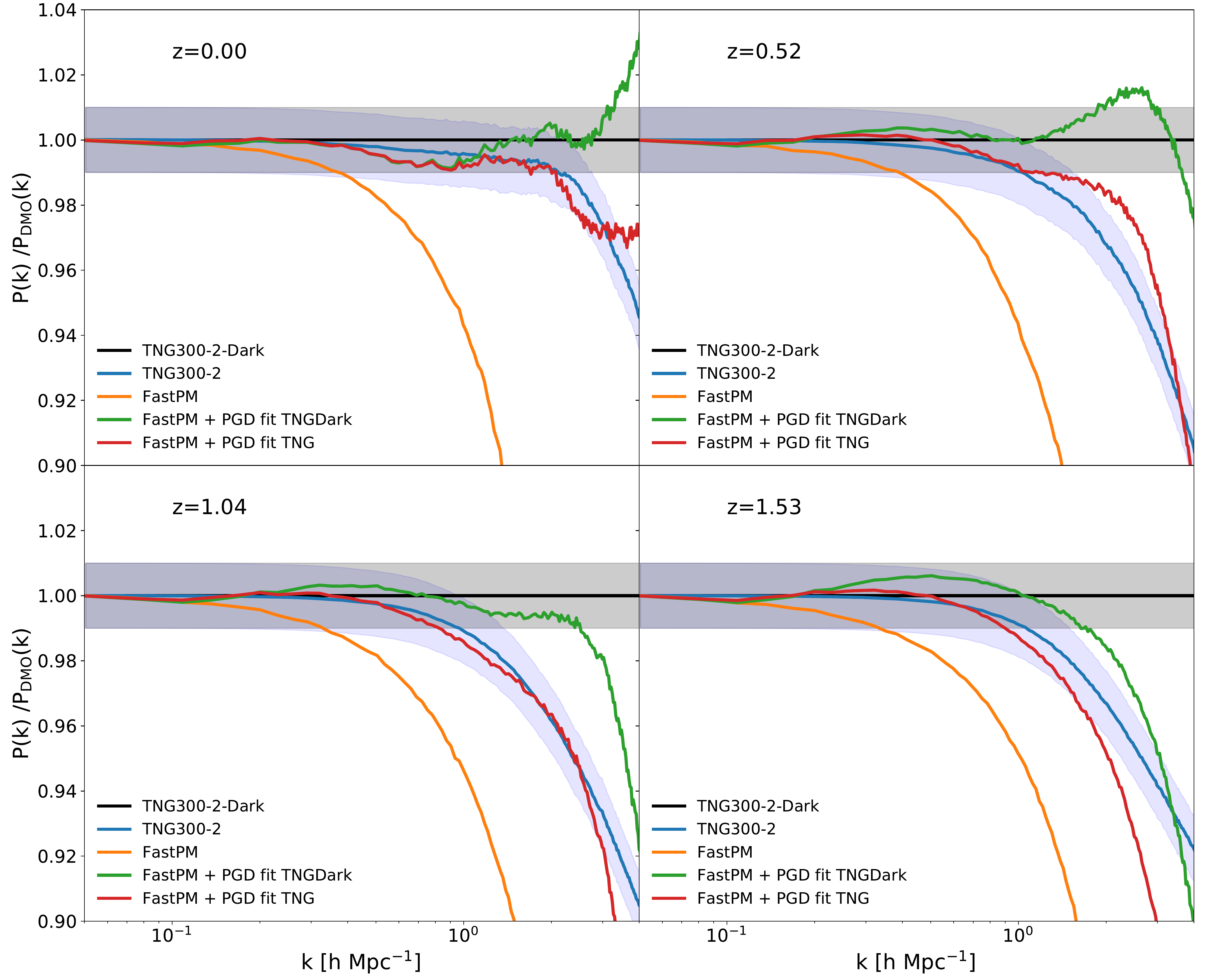}
\caption{The matter power spectrum of \FastPM{} simulation, before and after the calibration, and the reference simulation (TNG300 and TNG300-Dark) in different redshifts. Here we are trying to match the power spectrum of both TNG300-Dark and TNG300 (to account for the baryonic effect). The shadow region shows the $1\%$ deviation. The resolution of \FastPM{} is 125 times lower than TNG300-2-Dark.
}
\label{fig:PS_matter}
\end{figure}

The basic idea of the potential gradient descent (PGD) model is to add an additional displacement on the output of particles to mimic the missing sub-grid physics in \FastPM{} simulation. The additional displacement is modeled by the gradient of a modified gravitational potential, given by
\begin{align}
\mathbf{S} &= (\alpha/H_0^2)\ \mathbf{\nabla}\mathbf{\hat{O}_l\hat{O}_s}\phi\\
&= (4 \pi G \bar \rho\alpha/H_0^2)\ \mathbf{\nabla\hat{O}_l\hat{O}_s} \nabla^{-2} \delta
\label{eq:PGDPM}
\end{align}
where $\alpha$ is a free parameter, $\mathbf{\hat{O}_l}$ and $\mathbf{\hat{O}_s}$ are a high pass filter and a low pass filter, respectively
\be
\mathbf{\hat{O}_l}(k) = \exp{(-\frac{k_l^2}{k^2})}
\ee
\be
\mathbf{\hat{O}_s}(k) = \exp{(-\frac{k^4}{k_s^4})} 
\ee
Here $k_l$ and $k_s$ are also free parameters.

\cite{dai2018a} showed that the PGD model improves the halo profiles and the small scale power spectrum. The PGD model can be treated as a single post processing correction on the output snapshot, but it would be hard to do the correction on a lightcone output in this way, since the correction parameters are functions of redshift, and the redshift is not fixed in a lightcone output. Here we try to solve the problem by incorporating PGD correction into \FastPM{}. We perform a PGD correction after each \FastPM{} step, and then feed the corrected particle position into the next time step. Because the PGD is coupled into the simulation, both of the static snapshots and the lightcone output are consistently corrected.

The parameters $\alpha$ and $k_l$ are functions of redshift, we model their redshift dependence by
\be
\log(\frac{\alpha}{\alpha_0}) = Aa^2-Ba 
\ee
\be
k_l = k_{l,0}a^{\gamma}
\ee

where $\alpha_0$, $A$, $B$, $k_{l,0}$ and $\gamma$ are free parameters. $k_s = k_{s,0}$ is another free parameter and is fixed for all redshift. These parameters are fitted by matching the matter power spectrum at all redshift simultaneously. In figure \ref{fig:PS_matter} we show the matter power spectrum of original \FastPM{}, and \FastPM{} after the correction. Unlike \cite{dai2018a} where they are comparing the same resolution simulation, here the mass resolution of \FastPM{} is 125 times lower than the TNG300-2-Dark, yet we show that we can match the power spectrum quite well. The cross correlation coefficient also improves on all redshift, e.g., it improves approximately from 0.5 to 0.6 at the scale of $k=10 \hmpc$.

\subsection{Light-cone Simulation}

To test how the PGD model improves the weak lensing map, We build a light-cone output from the \FastPM{} simulation. The \FastPM{} simulation is run in a $3200 \hmpc$ periodic box with $1536^3$ dark matter particles. The simulation starts at redshift 9, with time steps separated by constant spacing in the scale factor. The \FastPM{} without PGD correction has 40 time steps, while after implementing the PGD correction we reduce the time steps to 20, since it has a comparable computation cost as a 40 step \FastPM{}.
The positions and velocities of the particles located between the steps are interpolated from nearest steps and are saved as the particle positions intersect the observer's light-cone. 
An optional FoF halo finder can be ran on the fly as the light-cone is generated, which uses padding to handle the continuity between light-cone slices.
Given that the volume of the simulation box can be smaller than the 
light-cone, the simulation box is tiled (duplicated) as necessary to cover the required volume of the lightcone. The light-cone module of \FastPM{} allows configurations on the position of the observer, the field of view angle ( determines the sky-area), the direction of sightlines, the replication (tiling) matrix, the list of culling octants, and the redshift range of interest.

In this work we assume the observer sits at the origin of the simulation box, and integrate the sightlines up to $z=2.2$. 
Note that the comoving distance to redshift 2.2 is around $3.8 \hgpc$, and the box is replicated along all directions to include the extra $600 \hmpc$.

\subsection{Weak Lensing Convergence}

\begin{figure}
\includegraphics[width=\textwidth]{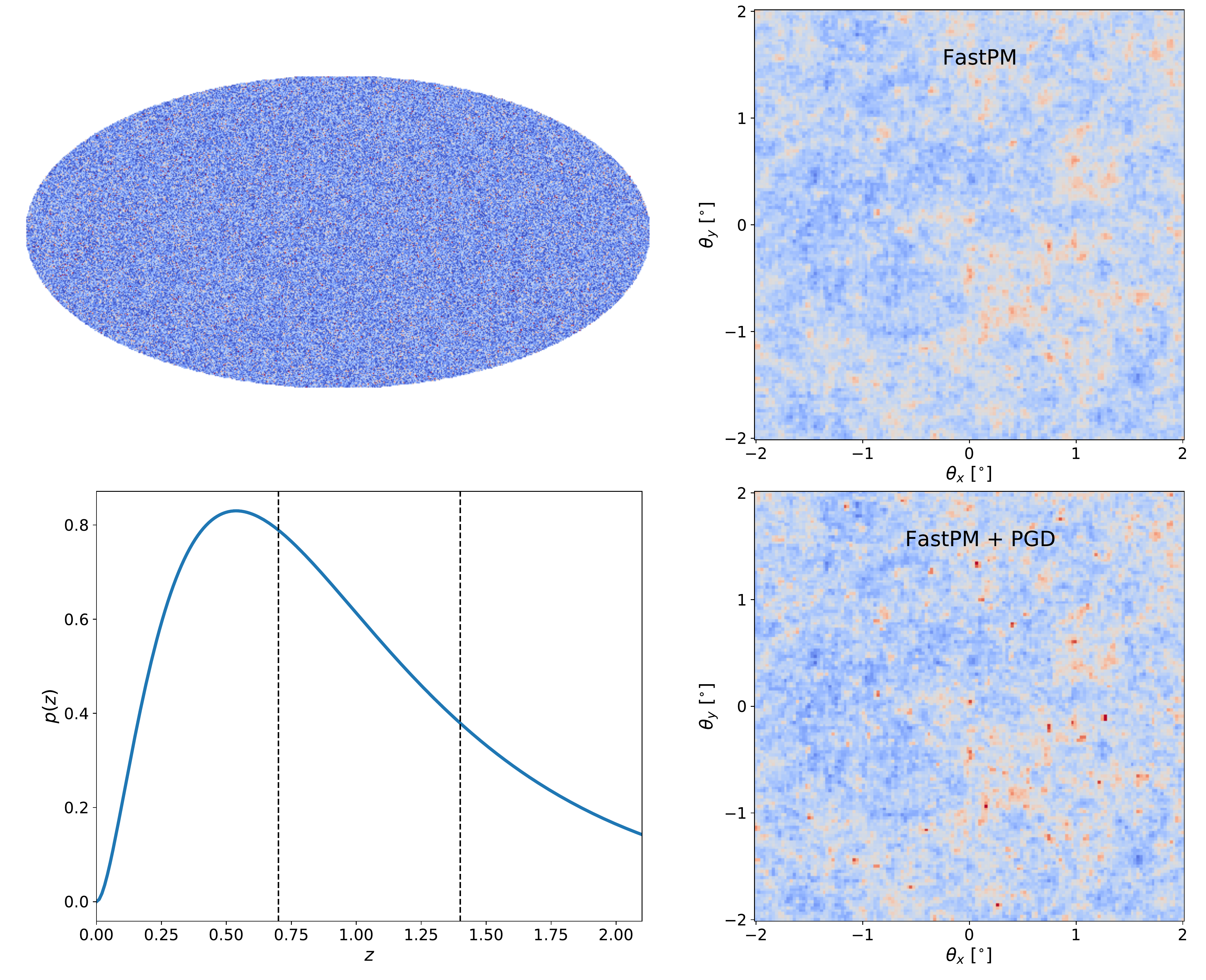}
\caption{The Mollweide view of weak lensing convergence field (upper left panel), the redshift distribution of source galaxies of a LSST-like survey (lower left panel), and the zoomed-in convergence field of \FastPM{} (upper right panel) and \FastPM{} with PGD correction (lower right panel).
\label{fig:weaklensing_kappa}}
\end{figure}

\begin{figure}
\includegraphics[width=\textwidth]{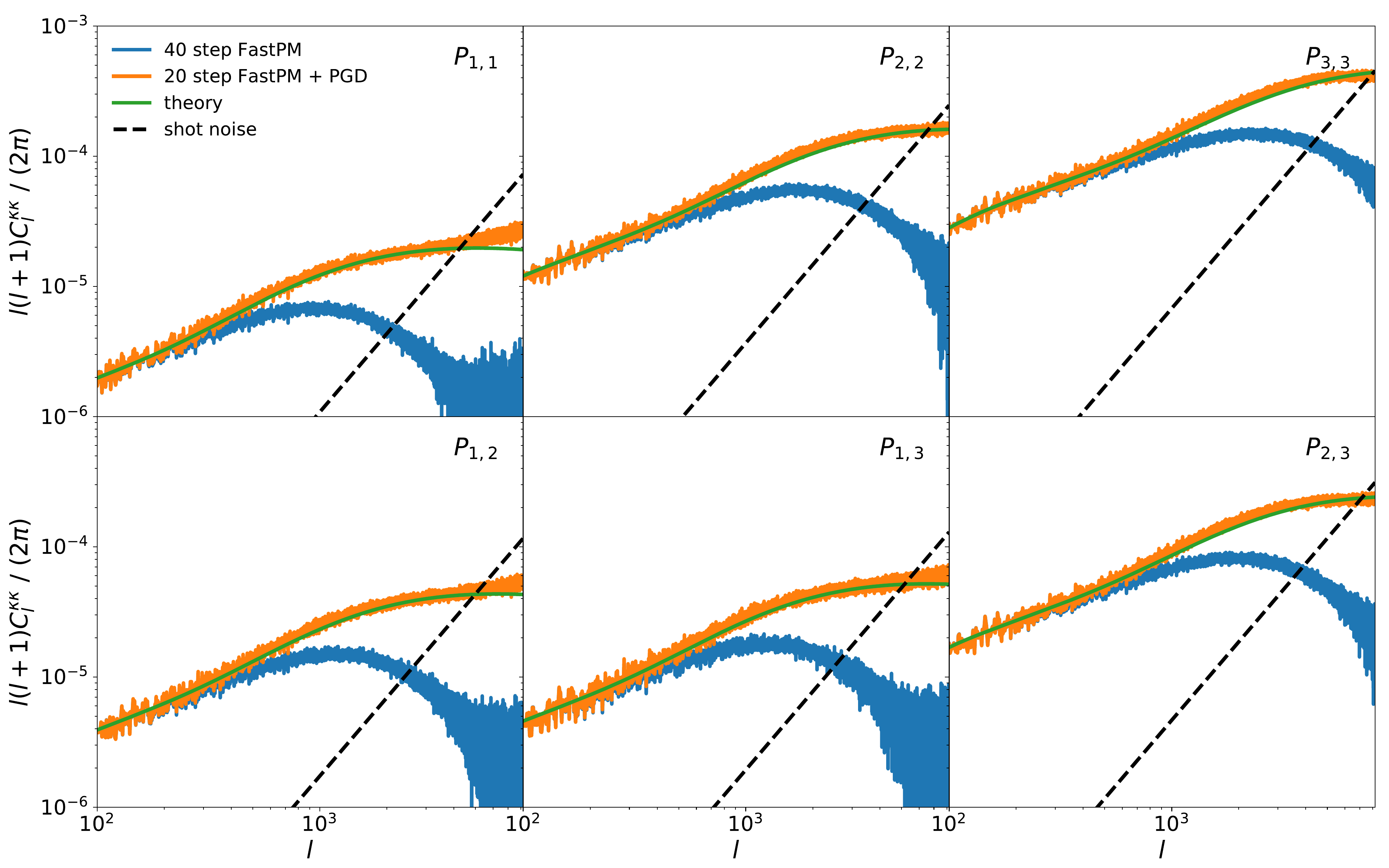}
\caption{The weak lensing convergence auto power spectrum (upper panel) and cross power spectrum (bottom panel). The bin number 1, 2 and 3 corresponding to tomographic bin $z\in[0, 0.7]$, $[0.7, 1.4]$ and $[1.4, 2.1]$, respectively. The theoretical weak lensing convergence power spectrum is calculated using Equation \ref{eq:Clkappa} with halofit nonlinear matter power spectrum.
Black dashed lines show the shot noise power spectrum from the particles, which we subtract from the measured power spectrum.
}
\label{fig:weaklensing}
\end{figure}

Under the Born approximation, we estimate the weak lensing convergence map produced by the source galaxies between redshift $z_{\mathrm{min}}$ and $z_{\mathrm{max}}$ as (see e.g. \cite{hoekstra2008a, kilbinger2015a})
\be
\kappa(\theta) = \frac{3H_0^2\Omega_m}{2c^2}\int_{z_{\rm min}}^{z_{\rm max}} p(z_s) dz_s \int_0^{\chi_s(z_s)} d\chi \frac{(\chi_s(z_s)-\chi)\chi}{\chi_s(z_s)a} \delta(\chi,\theta)
\ee
where $\theta$ is a 2D angular vector, $\delta$ is the matter overdensity at radial comoving distance $\chi$ and angular position $\theta$, $\chi_s(z_s)$ is the comoving distance to redshift $z_s$, and $p(z_s)$ is the normalized redshift distribution of source galaxies between redshift $z_{\mathrm{min}}$ and $z_{\mathrm{max}}$. Weak lensing maps are generated in the post processing after the simulation has ended and the particle lightcone has been saved. In the post processing step, the lightcone particles are read in, integrated along the line of sight using the weights based on lensing kernel and then a pixelized map is generated using HEALPY\citep{zonca2019a}, the python version of healpix \citep{gorski2005a}. While there is one I/O overhead due to lensing maps being generated in this manner, the post processing provides flexibility to generate multiple lensing maps for different lensing source configurations and saving the lightcone also allows one to generate lightcones for different probes in general (the line of sight integration kernel can be different from lensing kernel). Furthermore, for the case of cross correlations, it is also possible to generate the maps integrating over a narrow lens redshift range rather than over a complete redshift range from sources to observers, if necessary to reduce the I/O load.

In this work we assume a source galaxy redshift distribution of a LSST-like survey (second panel of Figure \ref{fig:weaklensing_kappa}). We divide the source into 3 tomographic bins: $z\in[0, 0.7]$, $[0.7, 1.4]$ and $[1.4, 2.1]$, and generate the convergence maps produced by these 3 source bins separately. In Figure \ref{fig:weaklensing_kappa} we show the all-sky convergence map as well as the zoomed-in maps of the last tomographic bin. We see that the PGD correction makes the peaks more evident. Thus we expect that the correction could help with non-Gaussian statistics such as peak statistics. Below we will examine the auto power spectrum and cross power spectrum of these convergence maps. 

Under the Limber approximation, the angular power spectrum of the weak lensing convergence can be written as
\begin{equation}
\begin{aligned}
\label{eq:Clkappa}
C^{\kappa}(l) = &(\frac{3H_0^2\Omega_m}{2c^2})^2 \int_{z_{\rm min1}}^{z_{\rm max1}} p_1(z_{s,1}) dz_{s,1} \int_{z_{\rm min2}}^{z_{\rm max2}} p_2(z_{s,2}) dz_{s,2} \\
&\int_0^{\chi_s({\rm min}(z_{s,1}, z_{s,2}))} d\chi (\frac{\chi_s(z_{s,1})-\chi}{\chi_s(z_{s,1})a}) (\frac{\chi_s(z_{s,2})-\chi}{\chi_s(z_{s,2})a}) P_m(k=\frac{l+0.5}{\chi}, z(\chi))
\end{aligned}
\end{equation}
where $P_m(k,z)$ is the 3D matter power spectrum. We have assumed that $p_1(z)$ and $p_2(z)$ are normalized. $p_1(z)$ and $p_2(z)$ will be the same in the case of auto power spectrum, and different for the cross power spectrum.
In Figure \ref{fig:weaklensing} we show the theoretical convergence power spectrum calculated using halofit nonlinear matter power spectrum \cite{takahashi2012a}, as well as the power spectrum we measure using the simulated convergence map. After the PGD correction the power spectrum matches the theoretical prediction better.
\section{Conclusions}
\label{sec:conclusion}

In this paper we try to improve halo statistics and small scale matter distribution in low resolution fast quasi N-body simulations. For halos, we introduce relaxed-FoF, a modification to the standard FoF algorithm so that the linking length is a function of the halo mass. For smaller halos, relaxed-FOF increase the linking length to enhance the identification of small halos, to improve agreement on the halo mass function, and to reduce the fraction of missed halos. We reject fake halos by reducing the linking length for the halos with large velocity dispersions. The rejection procedure removes fake halos found in the high density regions, and therefore improves the halo bias. We find that using a high resolution mesh for the 2LPT initial condition enhances the identification of small halos. We verify the 
results on several halo statistics, including halo bias, halo mass function, halo auto power spectrum in real space and in redshift space, cross correlation coefficient with the reference halo catalog, and halo-matter cross power spectrum. We find that our relaxed-FoF halo finder improves all of these. The ratio of missed halo and the halo catalog cross correlation coefficient suggest that our halo catalog from \FastPM{} is comparable to the halo catalog from a full N-body simulation of the same mass resolution, while our catalog has better large scale auto power spectrum in real space and redshift space, as well as better halo-matter cross power spectrum.

We also incorporate the potential gradient descent (PGD) method into \FastPM{} simulation to improve the matter distribution at nonlinear scales. We couple the PGD correction into the FastPM time steps. We show that the fully coupled PGD correction improves the matter power spectrum measured from static snapshot at all redshifts, just as the previously studied static PGD method \cite{dai2018a}. We build a light-cone simulation from a PGD-enabled \FastPM{} simulation, by interpolating the particles location between the steps. We show that the PGD correction significantly improves the convergence tomographic power spectrum measured from the light-cone output. 

We plan to use \FastPM{} for mock catalogs of both spectroscopic surveys such as DESI and photometric/weak lensing surveys such as LSST. The techniques we developed here will be useful to improve the halo and matter statistics in those simulations, thus enabling one to simulate the whole survey at the required mass resolution. For example, for DESI one needs to resolve halos down to $10^{11}M_{\odot}$ and to cover the entire survey one needs volumes in excess of $(3\hgpc)^3$, which can be achieved with $10^12$ particle \FastPM{} simulations, similar to the one that has recently been run \cite{modi2019a}. 

\textbf{Acknowledgements}
The majority of the computation were performed on NERSC computing facilities Edison and Cori, billed under the cosmosim and m3058 repository. National Energy Research Scientific Computing Center (NERSC) is a U.S. Department of Energy Office of Science User Facility operated under Contract No. DE-AC02-05CH11231. We thank Dylan Nelson and the IllustrisTNG team for kindly providing the linear power spectrum and random seed of the IllustrisTNG simulations.
This material is based upon work supported by the National Science Foundation under Grant Numbers 1814370 and NSF 1839217, and by NASA under Grant Number 80NSSC18K1274.

\appendix

\bibliographystyle{revtex}
\bibliography{reference}

\end{document}